\documentclass[aps,prb,floatfix,amsmath,amssymb,nofootinbib,twocolumn,superscriptaddress]{revtex4-1}
\usepackage[usenames,dvipsnames]{color}

\usepackage{graphicx,xcolor}
\usepackage{dcolumn}
\usepackage{bm}
\usepackage[normalem]{ulem}
\usepackage{graphics}
\usepackage[caption=false]{subfig}  
\usepackage[abs]{overpic}
\usepackage{relsize}  

\allowdisplaybreaks[1]

\def\bea{\begin{eqnarray}}
\def\eea{\end{eqnarray}}
\def\nn{\nonumber}
\def\vor{v}
\def\cd{c}
\def\dd{d}
\def\clock{6}

\newcommand{\red}[1]{{\color{red}{#1}}}
\newcommand{\blue}[1]{{\color{blue}{#1}}}
\newcommand{\green}[1]{{\color{green}{#1}}}

\begin{document}



\title{Molten Antiferromagnets in Two Dimensions}

\author{Itamar Shamai}
\affiliation{Department of Physics, Technion, Haifa 32000, Israel}

\author{Daniel Podolsky}
\affiliation{Department of Physics, Technion, Haifa 32000, Israel}
\affiliation{ITAMP, Harvard-Smithsonian Center for Astrophysics, Cambridge, Massachusetts 02138, USA}

\date{\today}

\begin{abstract}
We study crystal melting in two-dimensional antiferromagnets, by analyzing the statistical mechanics of the six-state clock model on a lattice in which defects (dislocations and disclinations) are allowed to appear.  We show that the elementary dislocations bind to fractional magnetic vortices. We compute the phase diagram by mapping the system into a Coulomb gas model.  Surprisingly, we find that in the limit of dominant magnetic interactions, antiferromagnetism can survive even in the hexatic and liquid phases. The ensuing molten antiferromagnets are topologically ordered and are characterized by spontaneous symmetry breaking of a non-local order parameter.
\end{abstract}

\maketitle


{\em Introduction --} The crystal structure of an antiferromagnet strongly affects its magnetic ordering.  For example, square lattices naturally accommodate N\'eel order, in which spins alternate in direction, whereas triangular lattices favor $\sqrt{3}\times\sqrt{3}$ order,  in which spins align at $120^\circ$ relative to their neighbors.  In geometrically frustrated lattices, magnetic ordering may be suppressed altogether, giving rise to spin liquid phases \cite{balents2010spin}.   In this setting it is natural to ask, How is crystal melting modified in antiferromagnetic systems?  Can new phases arise due to the interaction between magnetic and lattice fluctuations? And, Is solid lattice order a necessary condition for antiferromagnetism?

In three dimensions, melting is a first order phase transition; magnetic interactions are unlikely to change this. By contrast, in two dimensions (2D), crystal melting can occur in two successive continuous phase transitions, with the appearance of an intermediate hexatic phase \cite{kosterlitz1973,halperin1978,nelson1978,nelson1979,young1979}.  The transitions correspond to the unbinding of dislocations, followed by that of disclinations.  Then, an intricate interplay between crystal melting and magnetism may occur.

In this Letter, we explore this interplay in the context of a six-state clock model on a triangular lattice.  This model arises in the study of structural transitions of 2D ion crystals \cite{podolsky2016}.  We analyze this system by mapping it into a Coulomb gas description that treats magnetic vortices and lattice dislocations on an equal footing. We derive the renormalization group (RG) equations and solve them to obtain the phase diagram.   We neglect disclinations at first, assuming they are costly, and reinstate their effect later in the analysis.

\begin{figure}
\centering
\begin{overpic}[width = 0.5\textwidth]{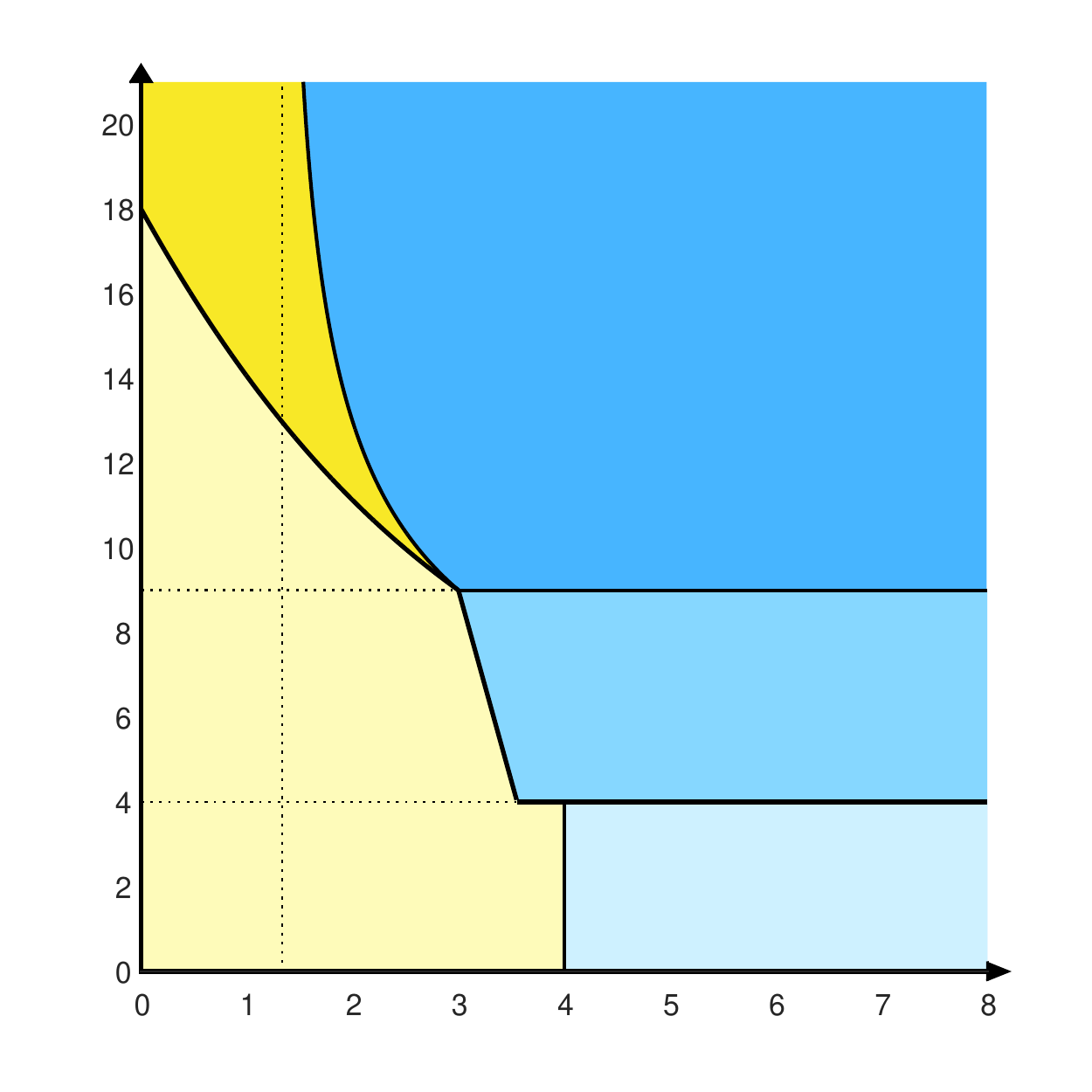}
  \put(0, 130){\large $K_m^0$}
  \put(130, 0){\large $K_l^0$}
  \put(163, 45.5){\large solid}
  \put(145, 88){\large Q-AF solid}
  \put(150, 154){\large AF solid}
  \put(53.7, 88){\large hexatic}
  \put(59, 202){\large $\longleftarrow\!$ \large AF hexatic}
  \put(133, 30.5){\footnotesize A}
  \put(133, 60){\footnotesize B}
  \put(232, 64.5){\footnotesize C}
  \put(115.5, 60.5){\footnotesize D}
  \put(107.5, 119.5){\footnotesize E}
  \put(232, 115.5){\footnotesize F}
  \put(68, 238){\footnotesize G}
  \put(34, 207){\footnotesize H}
  \put(63, 18.5){\tiny $\frac{4}{3}$}
\end{overpic}
\caption{Phase diagram for a system without disclinations, as a function of the bare reduced stifnesses $K_l^0$ and $K_m^0$. Each phase is labelled by its lattice ordering -- solid or hexatic -- and by its magnetic ordering -- antiferromagnetic (AF), quasi-long range antiferromagnetic (Q-AF), or disordered (no label).  The phase transitions are: Kosterlitz-Thouless (KT) transition driven by unbinding of dislocations (AB), KT transition of magnetic vortices (BC), cascaded transition (BD), KT transition of composite dislocations (DE), clock transition (EF), KT transition of double dislocations (EG) and a self-dual transition (EH).
}
\label{fig:phaseDiagClock}
\end{figure}
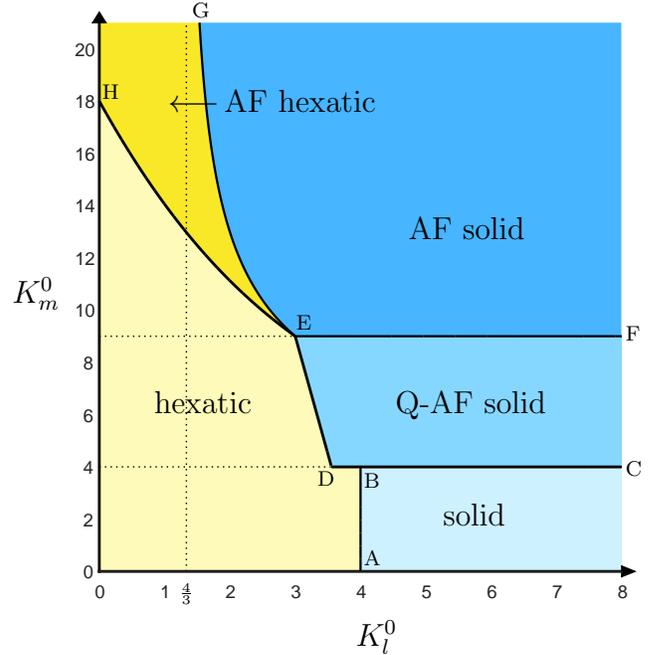


Figure~\ref{fig:phaseDiagClock} shows the phase diagram as a function of the two elastic energies characterizing our system.  The reduced (normalized by the temperature) magnetic stiffness, $K_m$,  measures the energy cost of a non-uniform magnetic configuration, while the reduced lattice stiffness, $K_{l}$, measures the Young modulus of the crystal.    Configurationally speaking, the system can be either a solid or a hexatic (when disclinations are added, a true liquid is also possible).  In 2D, a solid is characterized by algebraic translational order, long-range orientational order, and positive renormalized lattice stiffness $K_l>0$. A hexatic has short-range translational correlations, quasi-long range orientational order, and $K_l=0$.  Magnetically speaking, the system can have long range antiferromagnetic order (AF), with renormalized magnetic stiffness $K_m=\infty$; quasi-long range antiferromagnetic order (Q-AF), $0<K_m<\infty$; or be disordered, $K_m=0$.  Long range AF order is allowed because the clock term reduces the continuous $XY$ symmetry to a discrete one.

\begin{figure}
\subfloat[Single dislocation \label{fig:compositeDislocation}]{\begin{centering}
\begin{overpic}[width = 0.37\textwidth]{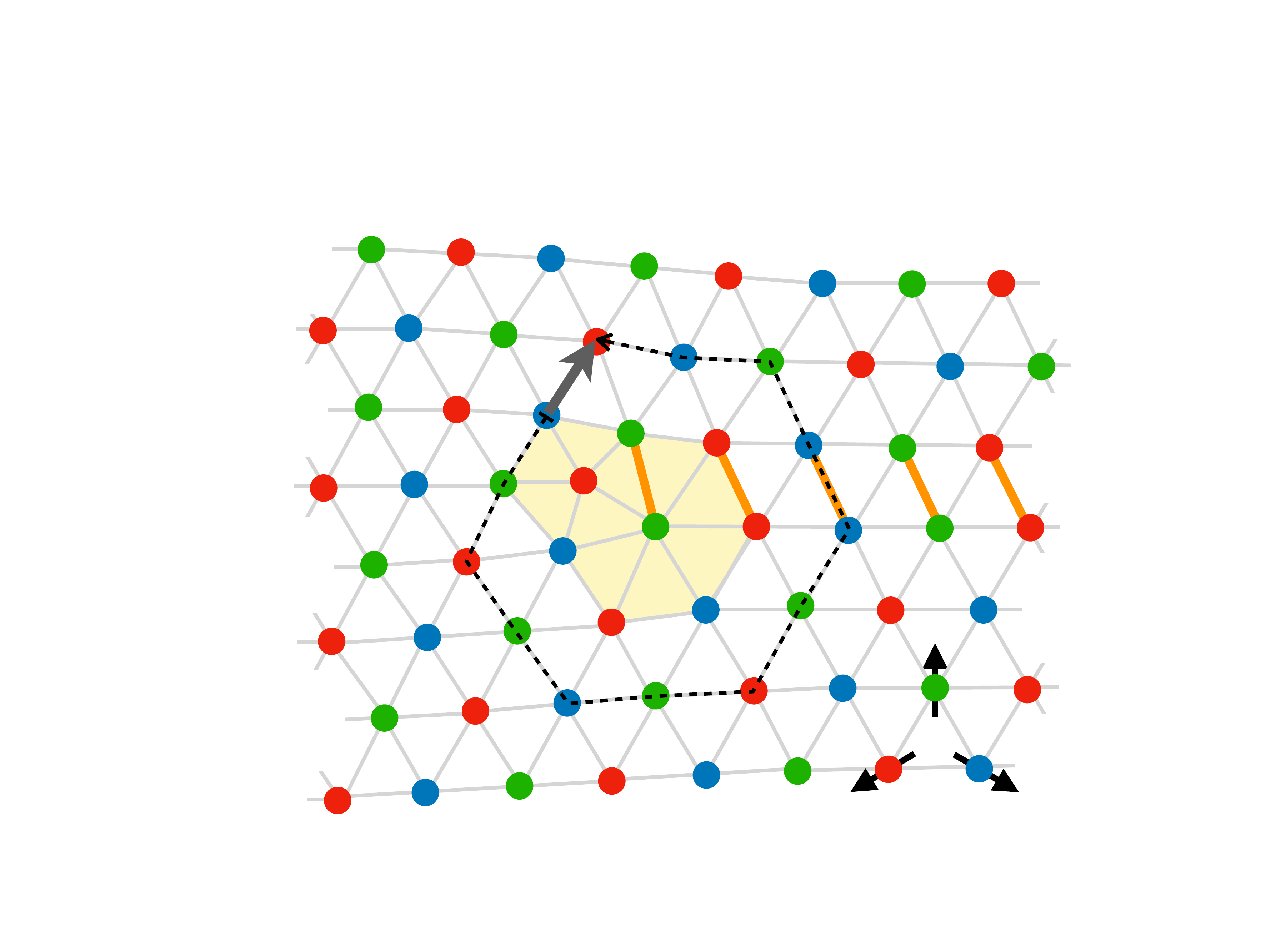}\end{overpic}
\end{centering}}
\hfill{}
\subfloat[Double dislocation \label{fig:doubleDislocation}]{\begin{centering}
\begin{overpic}[width = 0.36\textwidth]{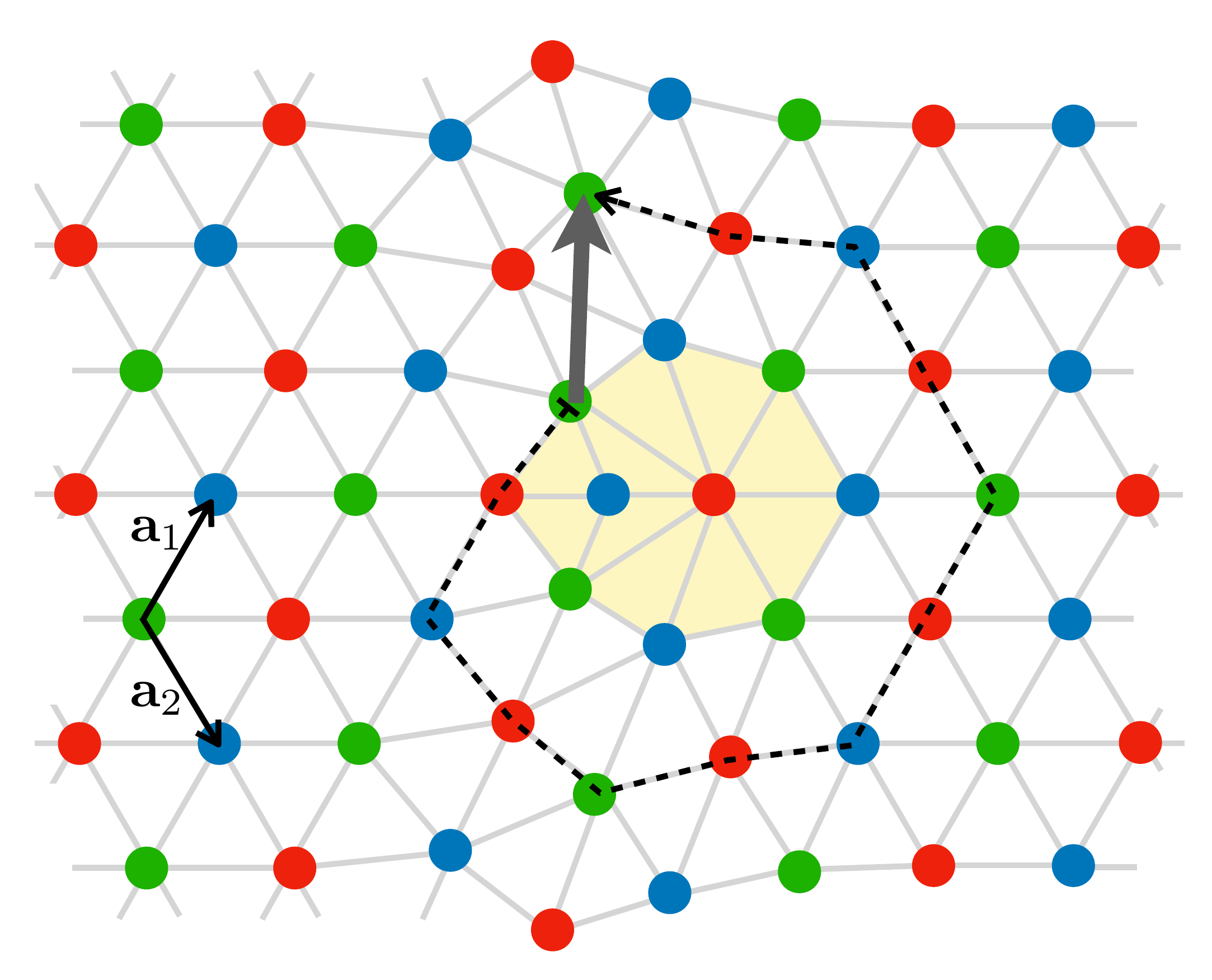}
\end{overpic}
\par\end{centering}}
 \caption{Dislocations of a triangular lattice.  For an $XY$ antiferromagnet, the colors represent spin direction, as shown in the lower-right corner of panel (a); for the buckled phase, they represent height.
(a) A dislocation of unit size Burgers' vector ${\bf b}={\bf a}_1$ (thick arrow) forms a domain wall, across which neighboring spins are aligned (orange bonds). The domain wall is removed by binding one third of a magnetic vortex (not shown).   (b) A double dislocation, with ${\bf b}={\bf a}_1-{\bf a}_2$, does not disrupt tripartite order.  Since $|{\bf b}|=\sqrt{3}a$, it costs three times as much lattice energy as a single dislocation.
\label{fig:dislocations} }
\end{figure}

{\em Antiferromagnetic hexatic --}  The most salient feature of Fig.~\ref{fig:phaseDiagClock} is the AF hexatic, a phase in which $K_m=\infty$ despite $K_l=0$.  It appears whenever
double dislocations, whose Burgers' vectors connect next-nearest-neighbors, proliferate before the single dislocations [see Fig.~\ref{fig:dislocations}].  This does not occur in ordinary cases of melting because double dislocations cost more energy.  However, as we show later, single dislocations in an antiferromagnet bind a third of a magnetic vortex.   These {\em composite dislocations}  cost both lattice and magnetic energy. Hence, for large enough $K_m^0$, they become more expensive than double dislocations, which do not bind magnetic vortices.  Double dislocations do not disrupt the tripartite structure needed for the AF order. Therefore, for any given snapshot of the molten lattice, it is possible to assign magnetic moments that maintain long-range  $\sqrt{3}\times\sqrt{3}$  AF order.

The AF hexatic is an unusual phase. From a strictly configurational point of view, it is a hexatic, with quasi-long range orientational order.  However, unlike a conventional hexatic, it is topologically ordered due to the lack of single dislocations.  From a magnetic point of view, it is characterized by long-range order of a non-local order parameter, involving a functional of the positions of all the particles, times the local spin. This follows since one can determine the relative direction of two distant spins, provided the spatial configuration of the particles in the intervening region separating the two spins is known. The situation is closely analogous to the zigzag phase of repulsive particles confined to 1D \cite{SMF1}.  There, the axial motion of the particles disorder the solid, and the zigzag is described by a string order parameter -- the relative orientation of two spins on the zigzag is determined only if the number of particles separating them is known \cite{dallaTorre2012}. 


The AF hexatic can be probed direcly through scattering experiments.  As shown in Appendix \ref{sec:scattering}, the magnetic structure factor of the hexatic has Lorentzian-shaped Bragg peaks, at the wave vectors {\bf K} associated with $\sqrt{3}\times\sqrt{3}$ ordering.  This signature distinguishes it clearly from the nonmagnetic hexatic.


{\em Composite dislocations -- } Consider an $XY$ antiferromagnet on a triangular lattice.  In the ground state, spins arrange in a $\sqrt{3}\times\sqrt{3}$ configuration.  However, as seen in Fig.~\ref{fig:compositeDislocation}, when a dislocation is added, it becomes impossible for the spins to follow this arrangement globally.  Instead, a domain wall is created, across which spins are aligned.  This costs energy that is linear in system size, making isolated dislocations prohibitive.
By binding to the dislocation a third of a magnetic vortex, the domain wall is removed.    Thus, {\em composite dislocations}, composed of a fractional 1/3 vortex bound to a dislocation, are elementary topological defects in this system \cite{timm2002}.

The composite dislocations encode the primary interplay between the lattice and the magnetism.  Binding of dislocations and disclinations to fractional vortices has been predicted in a large class of physical systems \cite{timm1998,timm2002,agterberg2011,RadzihovskyVishwanath,podolsky2009,berg2009,kruger2002,gopalakrishnan2013}.

{\em Buckled phase in ion crystals -- } Ions confined to 2D provide a physical realization of the six-state clock model. This system is known to undergo a series of structural transitions as the confining potential is relaxed \cite{dubin1,dubin2}.  For very strong confinement, the ions' repulsion results in a triangular lattice on a single plane.  When the confinement is relaxed, the ions undergo an instability, with a tendency to rearrange into a buckled configuration composed of three planes.  One third of the atoms stay in the base plane, one third are displaced above it, and one third below it, in a tripartite arrangement as shown in Fig.~\ref{fig:dislocations}.  In this system, we can think of the ion heights as effective magnetic degrees of freedom, and of the buckled phase as having AF order.

In Ref. \onlinecite{podolsky2016},  the buckling instability was mapped into a six state clock model using symmetry arguments.   An explicit mapping is provided in Appendix \ref{app:clock}. The six states correspond to the $3!$ choices of heights of the ions on a given reference triangular plaquette.  After coarse-graining, the long wave length description of this system is in terms of a {\em ferromagnetic} $XY$ model with an additional clock term, \cite{jose1977}
\bea
H_\mathrm{clock}=\int d^2 x\,\left[ \frac{\rho_s}{2} (\nabla\theta)^2+h_6\cos(6\theta)\right] \label{eq:clockHamiltonian}
\eea
However, the topological defects in this system remember that the fundamental degrees of freedom are repelling ion heights.  As shown in  Fig.~\ref{fig:compositeDislocation}, a dislocation gives rise to a domain wall, across which ions have matching heights.  As before, this is remedied by binding a third of a magnetic vortex.


{\em Coulomb gas model -- } We next write an effective model for the clock model in terms of low energy topological defects. This approach generalizes the Coulomb gas models used to describe the Kosterlitz-Thouless (KT) transition and two-dimensional melting \cite{kosterlitz1973}.

A topological defect is labelled by its magnetic charge, $q$; by its Burgers' vector\cite{chaikinLubensky}, $\mathbf{b}$; and by a dual clock charge, $n$ [see App.~\ref{app:coulombGas} for the definition of $n$]. We define for each defect $i$ a total charge $\mathbf{Q}_{i}=\left(q_{i},\mathbf{b}_{i},n_i\right)$.  The low energy topological defects are then divided into four classes: vortices, $\mathbf{Q}_{i}=\left(\pm1,0,0\right)$; composite dislocations, $\mathbf{Q}_{i}=\left(\pm\frac{1}{3},\mathbf{b},0\right),$ $\left|\mathbf{b}\right|=a$, where $a$ is the lattice constant; double dislocations, $\mathbf{Q}_{i}=\left(0,\mathbf{b},0\right),$ $\left|\mathbf{b}\right|=\sqrt{3}a$; and clock charges, $\mathbf{Q}_{i}=\left(0,0,\pm 1\right)$.

In the Coulomb gas representation, $K_{m}$ controls the strength of the interaction between vortices, whereas $K_{l}$ controls the interaction between dislocations. For the model in Eq.~(\ref{eq:clockHamiltonian}), $K_m=\frac{\rho_s}{T}$, whereas $K_{l}$ is given by the Young modulus of the crystal lattice.  Since the number of topological defects is not conserved, we assign a fugacity to each class of defect. We denote the fugacities of vortices, composite dislocations, double dislocations, and clock charges by $y_{\vor}$, $y_{c}$, $y_{\dd}$, and $y_6$, respectively.  The resulting Coulomb gas Hamiltonian is,
\begin{align}
-\beta H&=\sum_{i<j}\left(K_{m}q_{i}q_{j}+K_{l}\mathbf{b}_{i}\mathbf{\cdot b}_{j}
+\frac{6^{2}}{K_{m}}n_{i}n_{j}\right)\ln\frac{\left|\mathbf{r}_{i}\mathbf{-}\mathbf{r}_{j}\right|}{a}\nn\\
&\,+N_{\vor}\ln y_{\vor}+N_{\cd}\ln y_{\cd}+N_{\dd}\ln y_{\dd}+N_{\clock}\ln y_{\clock} \label{eq:CoulombGasEq}
\end{align}
where $N_{\vor,\cd,\dd,\clock}$ are the total number of defects in each class. See App.~\ref{app:coulombGas} for a derivation of this Hamiltonian.


{\em RG equations and phase diagram --} Following a renormalization group procedure for Coulomb gas models\cite{kosterlitz1973}, we obtain the flow equations,
\begin{align}
\frac{dK_{m}}{d\ell}&=-2\pi^{2}K_{m}^{2}\left(y_{\vor}^{2}+\frac{1}{3}y_{\cd}^{2}\right)+\frac{9}{2}\pi^{2}y_{\clock}^{2}\\
\frac{dK_{l}}{d\ell}&=-3\pi^{2}K_{l}^{2}\left(y_{\cd}^{2}+3y_{\dd}^{2}\right)\\
\frac{dy_{\vor}}{d\ell}&=\left(2-\frac{K_{m}}{2}\right)y_{\vor}\\
\frac{dy_{\cd}}{d\ell}&=\left(2-\frac{K_{l}}{2}-\frac{K_{m}}{18}\right)y_{\cd}\\
\frac{dy_{\dd}}{d\ell}&=\left(2-\frac{3}{2}K_{l}\right)y_{\dd}\\
\frac{dy_{\clock}}{d\ell}&=\left(2-\frac{18}{K_{m}}\right)y_{\clock} \label{eq:RGEqations}
\end{align}
These equations describe how the stiffness energies and the fugacities change as the system is coarse-grained, where $\ell$ is the coarse graining scale parameter.




Figure~\ref{fig:phaseDiagClock} shows the phase diagram obtained for our model after integrating the flow equations. The axes show the bare value of the lattice and magnetic stiffnesses, $K_l^0$ and $K_m^0$, assuming infinitesimal bare values of all the fugacities.  If the bare fugacities are small but not infinitesimal, then the exact position of the phase transition lines will shift, but the topology of the phase diagram and the universality class of the transitions will not be affected.

In order to understand the phase diagram, let's first consider a few simple limits. For $K_{m}^0<4$, the magnetic stiffness is too small to sustain magnetic ordering.  Then, the system undergoes a solid-to-hexatic melting transition through proliferation of dislocations, at the usual universal value $K_{l}^0=4$ (line AB on the phase diagram).  On the other hand, when $K_{l}^0$ is large, the lattice fluctuations are small and the system behaves as a six-state clock model on a perfect lattice \cite{jose1977}.  From the RG equations, we find two transitions: at $K_{m}^0=4$ (line BC) vortices unbind in a KT transition between the Q-AF solid and disordered solid; at $K_{m}=9$ (line EF), the clock term becomes relevant leading to a long-range AF solid.

The first unconventional elements in Fig.~\ref{fig:phaseDiagClock} are the direct transitions from Q-AF solid to the hexatic  (BD and DE lines).  At these transitions, the lattice and magnetic orders are simultaneously affected, yet the transitions are continuous.  Along DE, corresponding to $K_{l}+K_m/9=4$, the system undergoes a KT transition driven by unbinding of composite dislocations. Here, $K_l$ and $K_m$ do not separately acquire universal values.  Rather, only a specific combination of the two, $K_l+K_m/9$, is universal.   As a consequence, the power law correlations of the lattice and magnetic degrees of freedom do not individually have standard universal values.   Along BD, the system undergoes a cascaded transition, in which unbinding of vortices at an early stage of the RG flow eventually leads to unbinding of composite dislocations at a later stage.  This transition is also continuous, and is characterized by two diverging length scales \cite{podolsky2009}, with a lattice correlation length that is parametrically larger than the magnetic correlation length.

The transition from AF solid to AF hexatic (EG curve) involves the proliferation of double dislocations at a KT transition, with a universal renormalized lattice stiffness $K_l=4/3$. This is much less than the typical value at the solid-hexatic transition, $K_l=4$, and reflects the possibility to stabilize soft solids due to the magnetic suppression of single dislocations. The transition does not appear as a vertical line in Fig.~\ref{fig:phaseDiagClock}, due to strong renormalization of bare parameters, as described next.

The remainder of the phase diagram involves a competition between $y_c$ and $y_6$.  In the region above the line $K_{m}^0=9$, and to the left of the line $K_l^0+K_m^0/9=4$, both of these fugacities are initially relevant in the RG flow.  Along the curve $\frac{K_l^0}{2}+\frac{K_m^0}{18}=\frac{18}{K_m^0}$ both fugacities diverge at the same rate. This curve gives an estimate of the transition (EH) between AF hexatic and hexatic.  Below this curve, $y_c$ is dominant, hence renormalizing both $K_m$ and $K_l$ to zero.  This leads to the unbinding of composite dislocations and to the hexatic phase.  On the other hand, above this curve, $y_6$ dominates, thus renormalizing $K_m$ to infinity.  This, in turn, reverses the flow of $y_c$, leading to the eventual binding of composite dislocations.

We now consider two separate cases.  If one starts well above the $\frac{K_l^0}{2}+\frac{K_m^0}{18}=\frac{18}{K_m^0}$ curve, then the RG flow lead to the AF solid, as described above.   However, if one is only slightly above this curve, then $K_l$ decreases sufficiently during the RG flow that double dislocations become relevant.  Then, even though $y_c$ eventually vanishes due to the divergence in $K_m$, the system reaches the AF hexatic instead of the AF solid.  Hence, we conclude that there is no direct transition between the hexatic and the AF solid, but that a AF hexatic always intervenes between the two phases.  Near the point E, this intervening phase constitutes a thin sliver in the phase diagram.

Finally, the transition between the two hexatics (EH curve) is best understood in the vicinity of point $H$, where $K_l$ can be taken to be zero.  Since the transition occurs far from where the magnetic vortices become relevant ($K_{m}=4$), we neglect their effect on the magnetic stiffness. The reduced set of flow equations is then
$\frac{dK_{m}}{d\ell}=-2\pi^{2}K_{m}^{2}\frac{y_{c}^{2}}{3}+\frac{9}{2}\pi^{2}y_{6}^{2}$, $\frac{dy_{c}}{d\ell}=\left(2-\frac{K_{m}}{18}\right)y_{c}$, and $\frac{dy_{6}}{d\ell}=\left(2-\frac{18}{K_{m}}\right)y_{6}$.
This set of equations is invariant under the duality transformation
$\left(K_{m},4\sqrt{3}y_{c},y_{6}\right)\rightarrow\left(\frac{18^{2}}{K_{m}},y_{6},4\sqrt{3}y_{c}\right)$.
The point $K_{m}=18$ is a fixed point of the transformation provided that $\frac{y_{6}}{y_{c}}=4\sqrt{3}$.
Therefore, this is a self-dual transition and it is located at $K_{m}=18$ and $K_l=0$.  Note that this is a runaway flow, since at  $K_m=18$, both $y_{c}$ and $y_6$ diverge (at identical rates).  This may indicate a first order transition, although the self-duality suggests a continuous transition instead.  Moreover, the onset of topological order hints that the transition may be most naturally described by emergent gauge fields \cite{LammertRokhsarToner}.

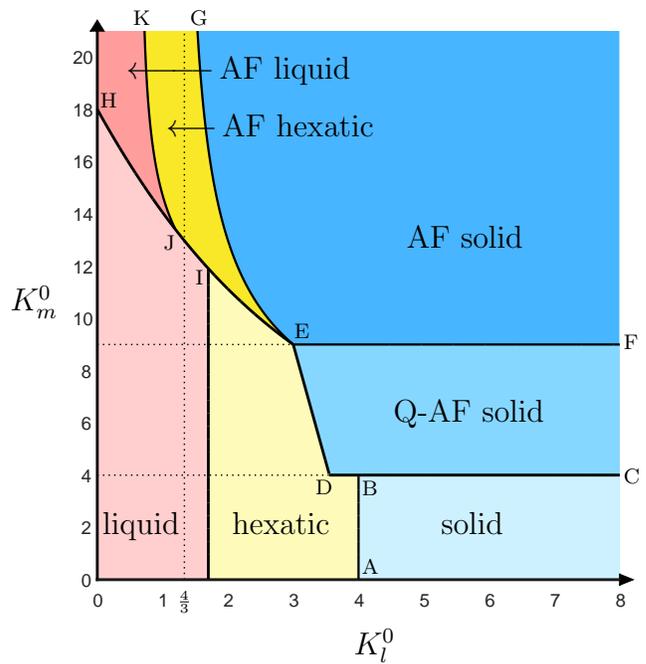
\begin{figure}
\centering
\begin{overpic}[width = 0.5\textwidth]{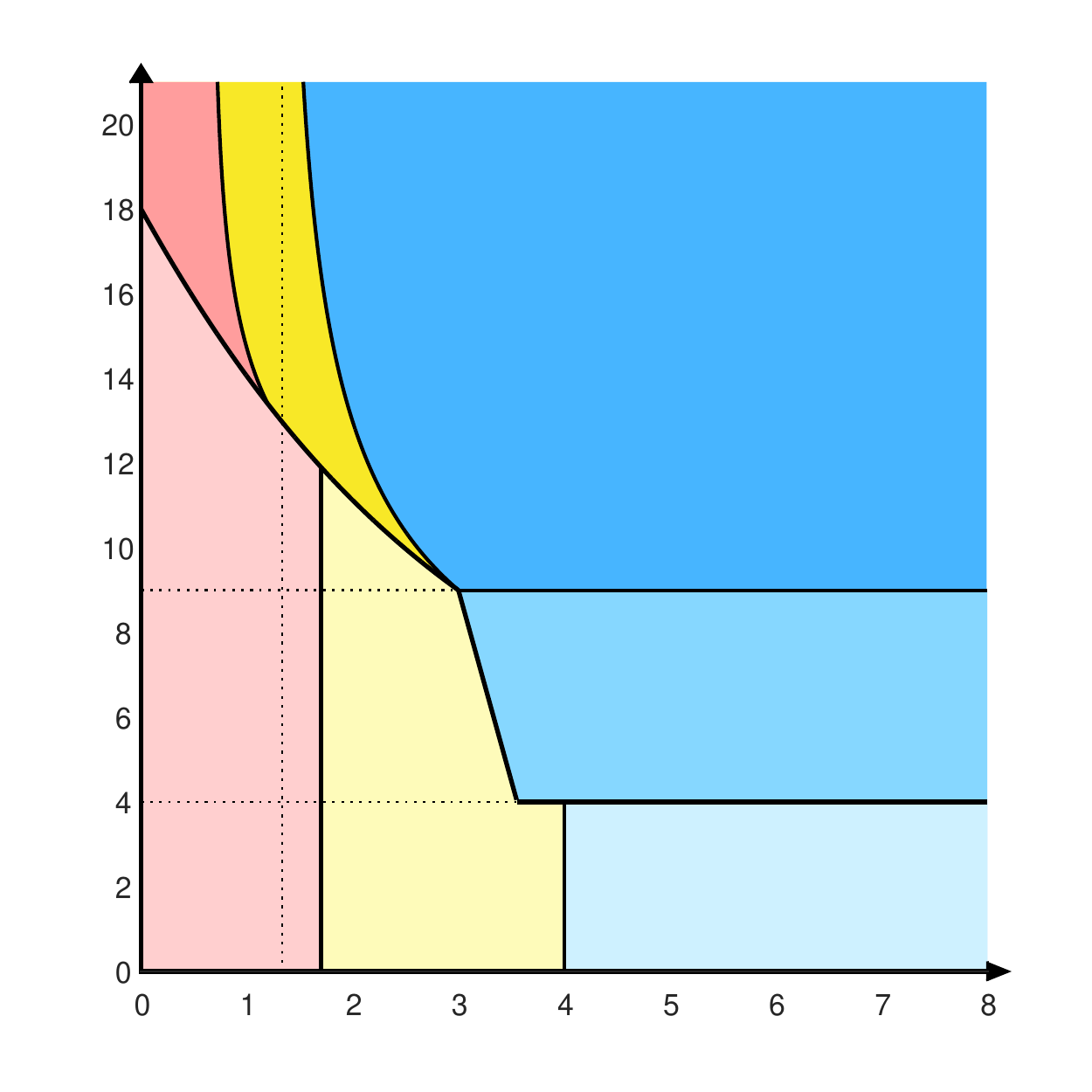}
  \put(0, 130){\large $K_m^0$}
  \put(130, 0){\large $K_l^0$}
  \put(163, 45.5){\large solid}
  \put(145, 88){\large Q-AF solid}
  \put(150, 154){\large AF solid}
  \put(59, 196){\large $\longleftarrow\!$ AF hexatic}
   \put(84, 45.5){\large hexatic}
   \put(35, 45.5){\large liquid}
   \put(44, 218){\large $\longleftarrow\!\!\!-\!-\!$ AF liquid}
  \put(133, 30.5){\footnotesize A}
  \put(133, 60){\footnotesize B}
  \put(232, 64.5){\footnotesize C}
  \put(115.5, 60.5){\footnotesize D}
  \put(107.5, 119.5){\footnotesize E}
  \put(232, 115.5){\footnotesize F}
  \put(34, 207){\footnotesize H}
  \put(68, 238){\footnotesize G}
  \put(70, 140){\footnotesize I}
  \put(58, 153){\footnotesize J}
  \put(46.5, 238){\footnotesize K}
  \put(63, 18.5){\tiny $\frac{4}{3}$}
\end{overpic}
\caption{Phase diagram with disclinations taken into account. There are a number of possible scenarios -- here we display the case where disclinations are costly relative to the dislocations. Otherwise, the hexatic phases may be preempted by direct first order transitions from the solids to the liquids.}
\label{fig:phaseDiagramDiscl}
\end{figure}

{\em Antiferromagnetic liquid -- } Thus far we have neglected the effect of disclinations. As explained in App.~\ref{SI:disclinations}, disclinations cost magnetic energy that is linear in system size. However, unlike dislocations, disclinations cannot reduce their energy by binding a fractional vortex. Therefore, disclinations are suppressed in regions where $K_m$ is large. By contrast, double disclinations are compatible with magnetic ordering; hence, they can unbind even when the magnetism is relevant.  This raises the possibility of a AF liquid phase, in which double dislocations and disclinations unbind while single dislocations and disclinations are bound.  Figure \ref{fig:phaseDiagramDiscl} shows a possible phase diagram. The AF hexatic is no longer guaranteed to exist in this case, since it may be precluded by a direct first-order melting transition from the AF solid into the AF liquid. However, for sufficiently large $K_m$ and small $K_l$, the AF liquid is guaranteed to appear.

{\em Discussion -- } The possibility of an antiferromagnetic liquid has been raised before for crystals with many-site bases, which naturally accommodate antiferromagnetic order within the unit cell \cite{timm2002}.  Then, the antiferromagnetism does not break the translational symmetry of the lattice and therefore dislocations do not frustrate the magnetic order. By contrast, in our model, the crystal is a Bravais lattice, the antiferromagnetism occurs at a non-zero wave vector ${\bf K}$, and the antiferromagnetic correlations are protected dynamically.

It is interesting to consider how this analysis generalizes to other closely related systems. Appendix \ref{app:generalizations} shows two such generalizations, to the four-state clock model and the XY model, both on the square lattice. For the four-state clock model, we find that the AF hexatic can be reached if the magnetic interaction is slightly larger than the lattice interaction, making the phase potentially more accessible to experiment. For the $XY$ model on the square lattice, true long-range order is not possible, but Q-AF hexatic and Q-AF liquid phases appear for large enough $K_m/K_l$.

\acknowledgements
We are grateful to Ehud Altman, Shmuel Fishman, Giovanna Morigi, Achim Rosch, Efrat Shimshoni, and Ashvin Vishwanath for useful discussions.  We thank support by the Israel Science Foundation (ISF)
grant numbers 1839/13, by the Joint UGS-ISF Research Grant Program under grant number 1903/14, by the National Science Foundation through a grant to ITAMP at the Harvard-Smithsonian Center for Astrophysics, and through support from the Harvard-MIT CUA.

\appendix

\section{Signatures of the Phases in Scattering Experiments}
\label{sec:scattering}

A direct way to detect experimentally the phases in Fig.~\ref{fig:phaseDiagClock} is by Bragg scattering experiments. We distinguish between scattering experiments that are sensitive only to the lattice (Bragg scattering) and those that are sensitive also to the magnetic degrees of freedom (magnetic Bragg scattering). Here, we derive the signatures of all the phases both for magnetic and non-magnetic cases. Each phase can be uniquely identified by the two kinds of signatures.

The structure factor for Bragg scattering is defined as,
\bea
S\left(\mathbf{k}\right)=\frac{1}{N}\sum_{ij}\left\langle e^{i\mathbf{k}\cdot\left(\mathbf{r}_{i}-\mathbf{r}_{j}\right)}\right\rangle
\label{eq:StructureFactorDefinition}
\eea
where $N$ is the total number of particles, and for the magnetic Bragg scattering it is,
\bea
S_{M}\left(\mathbf{k}\right)=\frac{1}{N}\sum_{ij}\left\langle e^{i\mathbf{k}\cdot\left(\mathbf{r}_{i}-\mathbf{r}_{j}\right)}e^{i\left(\theta\left(\mathbf{R}_{i}\right)-\theta\left(\mathbf{R}_{j}\right)\right)}\right\rangle
\label{eq:MagneticStructureFactorDefinition}
\eea
where the $\mathbf{r}_{i}$ are the positions of the ions, $\mathbf{R}_{i}$ are the equilibrium positions and $\theta\left(\mathbf{R}_{i}\right)$ are the magnetic degrees of freedom.

In order to evaluate these quantities, we write the position of each particle as $\mathbf{r}_{i}=\mathbf{R}_{i}+\mathbf{u}_{i}$, where $\mathbf{u}_i$ is the fluctuation about equilibrium.  In addition, when evaluating $S_M(\mathbf{k})$ we assume that the lattice and magnetic fluctuations decouple, and factorize Eq.(\ref{eq:MagneticStructureFactorDefinition}),
\bea
S_{M}\left(\mathbf{k}\right)=\frac{1}{N}\sum_{ij}\left\langle e^{i\mathbf{k}\cdot\left(\mathbf{u}_{i}-\mathbf{u}_{j}\right)}\right\rangle \left\langle e^{i\left(\theta\left(\mathbf{R}_{i}\right)-\theta\left(\mathbf{R}_{j}\right)\right)}\right\rangle e^{i\mathbf{k}\cdot\left(\mathbf{R}_{i}-\mathbf{R}_{j}\right)}\nonumber
 \eea
By translational invariance of the correlation functions we can write
\bea
S_{M}\left(\mathbf{k}\right)=\sum_{\mathbf{R}}A_{k}\left(\mathbf{R}\right)B\left(\mathbf{R}\right)e^{i\mathbf{k}\cdot\mathbf{R}}
\eea
where
\bea
A_{k}\left(\mathbf{R}_{ij}\right)\equiv\left\langle e^{i\mathbf{k}\cdot\left(\mathbf{u}_{i}-\mathbf{u}_{j}\right)}\right\rangle
\eea
\bea
B\left(\mathbf{R}_{ij}\right)\equiv\left\langle e^{i\left(\theta\left(\mathbf{R}_{i}\right)-\theta\left(\mathbf{R}_{j}\right)\right)}\right\rangle
\eea
We shift to Fourier space by a series of equalities:
\bea
S_{M}\left(\mathbf{k}\right)&=&\int d^{2} x\, A_{k}\left(\mathbf{x}\right)B\left(\mathbf{x}\right)e^{i\mathbf{k}\cdot\mathbf{x}}\sum_{\mathbf{R}}\delta\left(\mathbf{x}-\mathbf{R}\right)\\
&=&\sum_{\mathbf{G}}\int d^{2}x\, A_{k}\left(\mathbf{x}\right)B\left(\mathbf{x}\right)e^{i\mathbf{\left(k-G\right)}\cdot\mathbf{x}}\\
&=&\sum_{\mathbf{G}}\int d^2q\,\tilde{A}_{k}\left(\mathbf{q}\right)\tilde{B}\left(\mathbf{k}-\mathbf{G}-\mathbf{q}\right) \label{eq:MagneticStructureFactor}
\eea
where in the second step we used the Poisson summation formula, and where the wave vectors $\mathbf{G}$ are defined through the condition $\mathbf{G}\cdot\mathbf{R}=2\pi n$, $n\in\mathbb{Z}$.
The non-magnetic Bragg structure factor is obtained by setting $B\left(\mathbf{R}_{ij}\right)=1$ or equivalently $\tilde{B}\left(\mathbf{k}-\mathbf{G}-\mathbf{q}\right)=\delta\left(\mathbf{k}-\mathbf{G}-\mathbf{q}\right)$.
Our general expression for the Bragg structure factor is then
\bea
S\left(\mathbf{k}\right)=\sum_{\mathbf{G}}\tilde{A}_{k}\left(\mathbf{k}-\mathbf{G}\right),
\eea
and for magnetic Bragg structure factor
\bea
S_{M}\left(\mathbf{k}\right)=\sum_{\mathbf{G}}\int d^{2}\mathbf{q}\tilde{A}_{k}\left(\mathbf{q}\right)\tilde{B}\left(\mathbf{k}-\mathbf{G}-\mathbf{q}\right).
\eea
We now substitute into the general expressions the specific forms of $\tilde{A}_{k}\left(\mathbf{q}\right)$ and $\tilde{B}_{k}\left(\mathbf{q}\right)$ for each phase.

{\em Solid -- } For the two-dimensional solid phase, the fluctuations in the mean positions of the ions are Gaussian and scale logarithmically with distance:
\bea
A_{k}\left(\mathbf{R}\right)=e^{-\frac{k^{2}}{2}\left\langle \left(\mathbf{u}\left(\mathbf{R}\right)-\mathbf{u}\left(\mathbf{0}\right)\right)^{2}\right\rangle }\sim|\mathbf{R}|^{-\alpha k^{2}a_{0}^{2}}
\eea
where $\alpha$ is related to the renormalized Lam\'e elasticity coefficients of the lattice, and the temperature \cite{nelson1979} by:
\bea
\alpha=\frac{3\mu_{R}+\lambda_{R}}{4\pi\mu_{R}\left(2\mu_{R}+\lambda_{R}\right)}T
\eea
After Fourier-transforming, $\tilde{A}_{k}\left(\mathbf{q}\right)=\frac{1}{\mathbf{q}^{2-\alpha k^{2}a_{0}^{2}}}$, and the Bragg structure factor is a sum of power law singularities at the Bragg points $\mathbf{G}$:
\bea
S\left(\mathbf{k}\right)&\simeq&\sum_{\mathbf{G}}\frac{1}{\mathbf{\left|\mathbf{k}-\mathbf{G}\right|}^{2-\alpha k^{2}a_{0}^{2}}}\\
&\simeq&\sum_{\mathbf{G}}\frac{1}{\mathbf{\left|\mathbf{k}-\mathbf{G}\right|}^{2-\alpha G^{2}a_{0}^{2}}}
\eea

{\em AF solid --} For the AF solid, there is an additional magnetic signature
\bea
\tilde{B}\left(\mathbf{q}\right)=\sum_{\mathbf{K}}\delta\left(\mathbf{q}-\mathbf{K}\right)
\eea
where $\mathbf{K}$ are the wave vectors associated with the magnetism, and
\bea
S_{M}\left(\mathbf{k}\right)=\sum_{\mathbf{G},\mathbf{K}}\frac{1}{\mathbf{\left|\mathbf{k}-\mathbf{G}-\mathbf{K}\right|}^{2-\alpha\left(\mathbf{G}+\mathbf{K}\right)^{2}a_{0}^{2}}}
\eea
Note that the peaks in the magnetic scattering occur at wave vectors shifted by $\mathbf{K}$ relative to the non-magnetic Bragg points.
For the Q-AF solid, the magnetic correlations decay algebraically
\bea
\tilde{B}\left(\mathbf{q}\right)=\sum_{\mathbf{K}}\frac{1}{\mathbf{\left|\mathbf{q}-K\right|}^{2-\eta}}
\eea
where $\frac{1}{9}\leq\eta\leq\frac{1}{4}$. This results in broadening of peaks of the magnetic structure factor by $\eta$, compared to the previous case:
\bea
S_{M}\left(\mathbf{k}\right)&=&\sum_{\mathbf{G},\mathbf{K}}\int d^2 q\,\frac{1}{\mathbf{q}^{2-\alpha k^{2}a_{0}^{2}}}\frac{1}{\mathbf{\left|\mathbf{k}-\mathbf{G}-\mathbf{K}-q\right|}^{2-\eta}}\nonumber \\
&\simeq&\sum_{\mathbf{G},\mathbf{K}}\frac{1}{\mathbf{\left|\mathbf{k}-\mathbf{G}-\mathbf{K}\right|}^{2-\alpha k^{2}a_{0}^{2}-\eta}}\nonumber\\
&\simeq&\sum_{\mathbf{G},\mathbf{K}}\frac{1}{\mathbf{\left|\mathbf{k}-\mathbf{G}-\mathbf{K}\right|}^{2-\alpha\left(\mathbf{G}+\mathbf{K}\right)^{2}a_{0}^{2}-\eta}}
\eea

{\em Hexatic --} For the hexatic phase, the shape of the Bragg peaks are known to be Lorentzians \cite{davey1984}, and the stucture factor is
\bea
S\left(\mathbf{k}\right)=\sum_{\mathbf{G}}\frac{1}{\mathbf{\left|\mathbf{k}-\mathbf{G}\right|}^{2}+\gamma^{2}}\label{eq:LorentzianHexatic}
\eea
with $\gamma$ being the width of the Lorentzian.

{\em AF hexatic --} The AF hexatic's structure factor is of the same form as that of the hexatic, Eq.~(\ref{eq:LorentzianHexatic}).  Using
\bea
\tilde{B}\left(\mathbf{q}\right)=\sum_{\mathbf{K}}\delta\left(\mathbf{q}-\mathbf{K}\right)
\eea
we obtain the magnetic structure factor
\bea
S_{M}\left(\mathbf{k}\right)=\sum_{\mathbf{G},\mathbf{K}}\frac{1}{\mathbf{\left|\mathbf{k}-\mathbf{G}-\mathbf{K}\right|}^{2}+\gamma^{2}}.
\eea

\section{Mapping to six-state clock model}
\label{app:clock}

Consider a system of ions confined to a plane by a harmonic trap. Upon cooling, the ions crystalize into a solid on a triangular lattice.  When the confinement potential is relaxed, they undergo an instability into a buckled phase, in which the planar positions ${\bf r}=(x,y)$ remain unchanged, but the heights $z$ of the ions buckle in order to reduce their mutual repulsion.  In the buckled phase, the heights form a staggered pattern on a tripartite lattice, in which ions on one sublattice rise, $z>0$, on another they submerge, $z<0$, and on the third they remain level, $z=0$.  This pattern is captured by writing the height $z_i$ of the particle at planar location ${\bf r}_i$ as,
\begin{eqnarray}
z_i={\mathrm Re} \left[\psi e^{i{\mathbf K}\cdot{\mathbf r}_i}\right]
\end{eqnarray}
where ${\mathbf K}$ is the wave vector at the corner of the first Brillouin zone of the triangular lattice.  The variable $\psi=|\psi|e^{i\theta}$ is a complex number that acts as the order parameter.

In Ref. \onlinecite{podolsky2016}, it was shown that the buckled transition can be mapped into a six-state clock model.  This was done by expanding the Ginzburg-Landau free energy in powers of the order parameter $\psi$.  When the symmetries of the triangular lattice are taken into account, it was shown that in addition to terms such as $|\nabla\psi|^2$, $|\psi|^2$, $|\psi|^4$, and $|\psi|^6$ appearing in the free energy, the term $\frac{1}{2}\left[\psi^6+(\psi^*)^6\right]$ is also allowed.  This gives rise to a clock term $\cos(6\theta)$, which tends to pin  $\theta$ to one of six discrete values, corresponding to the $3!$ different choices of the height pattern.

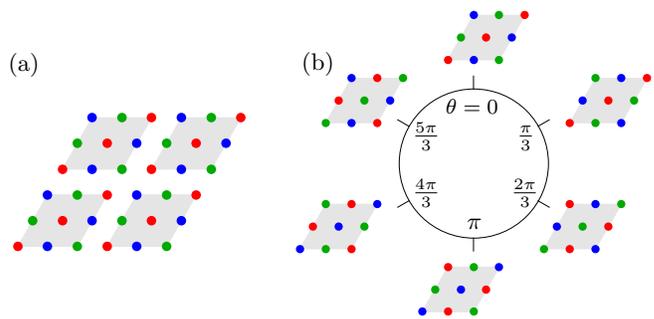
\begin{figure}
\centering
 \raisebox{9mm}{\begin{overpic}[width=0.18\textwidth]{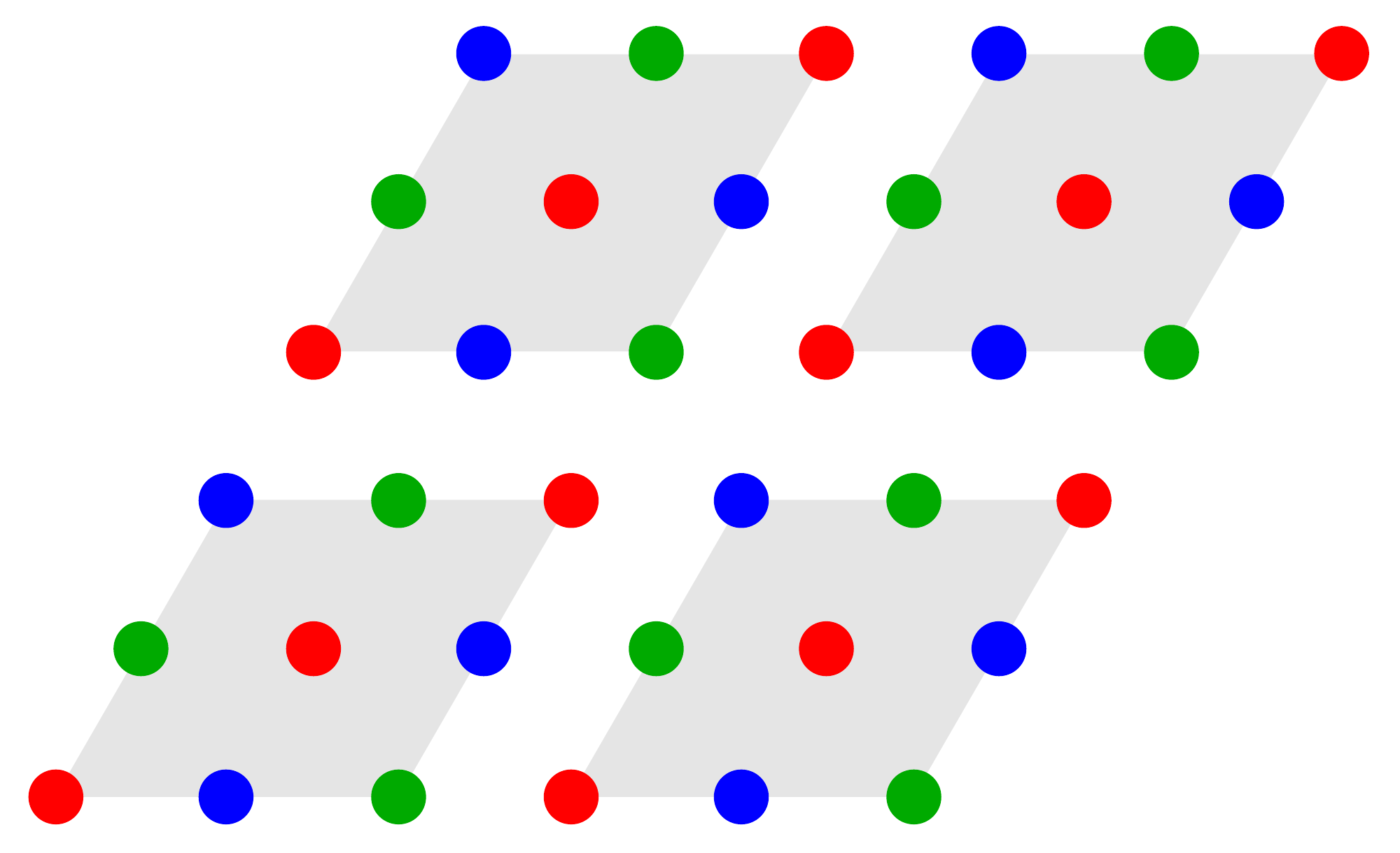}
  \put(0,70){(a)}
 \end{overpic}}
\hfill
\begin{overpic}[width = 0.275\textwidth]{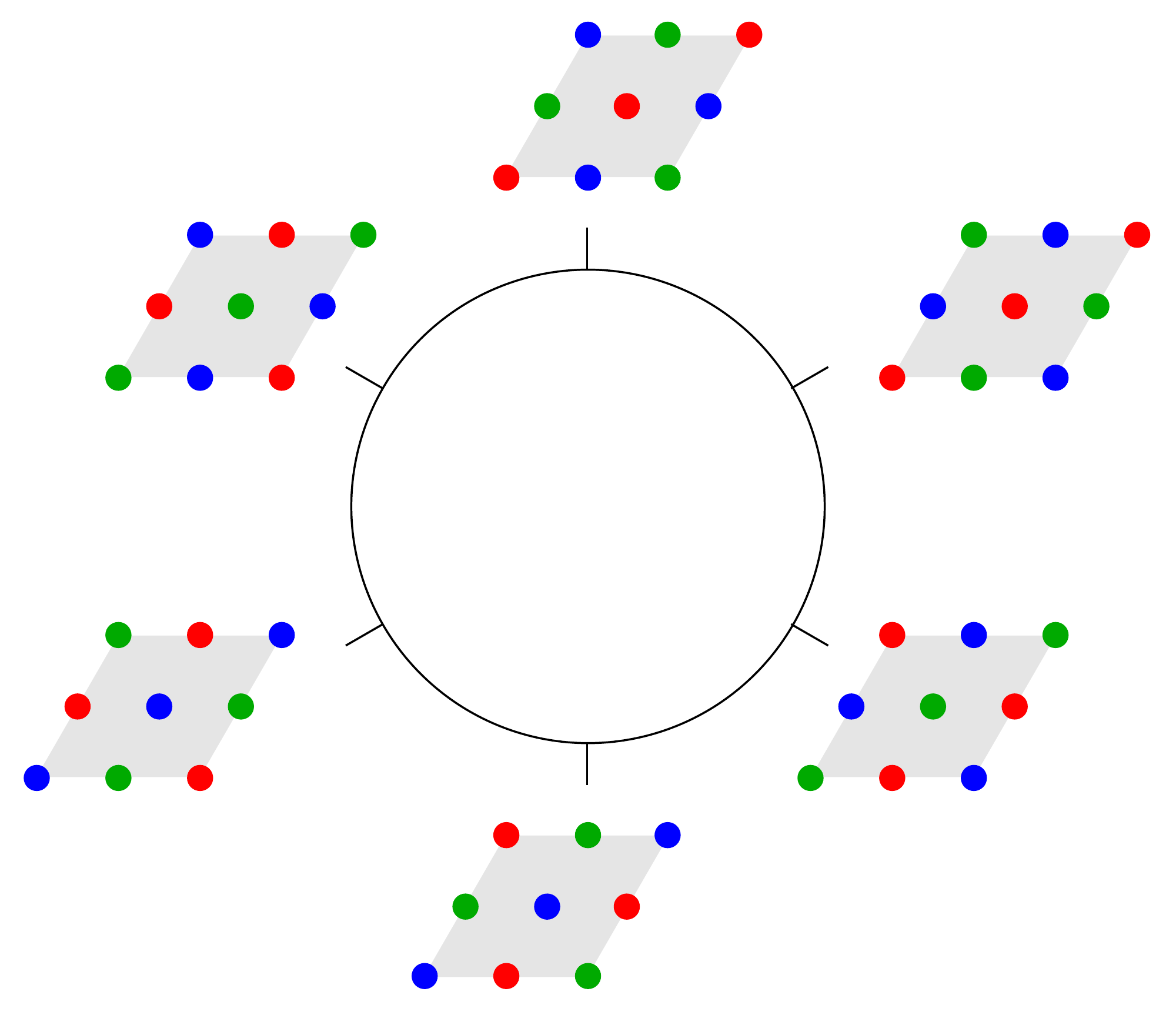}
 \put(5,96){(b)}
  \put(59.6, 79.4){\footnotesize $\theta=0$}
  \put(86.3, 68.8){$\frac{\pi}{3}$}
  \put(84.1, 47.3){$\frac{2\pi}{3}$}
  \put(67.6, 36){$\pi$}
  \put(46.8, 47.3){$\frac{4\pi}{3}$}
  \put(46.8, 68.8){$\frac{5\pi}{3}$}
\end{overpic}
\caption{(a) Coarse graining of the buckled phase on a $3\times 3$ plaquete, corresponding to N=1.  The color of the ions represents their height. (b) Clock mapping.}
\label{fig:clock mapping}
\end{figure}

Here, the mapping into a 6-state clock model is carried out explicitly, as follows.  We coarse-grain the system in patches of $3N\times 3N$ ions, as shown in Fig. \ref{fig:clock mapping}(a). Since there are $3!$ possible height patterns, we assign to each patch a discrete angle  $\theta=\frac{2\pi n}{6}$, as shown in Fig. \ref{fig:clock mapping}(b). We then compute the domain wall energy between pairs of patches, arising from the Coulomb interaction between the ions composing these patches.  We find that, for large $N$, the domain wall energies are independent of the relative orientation of the patches.  The domain wall energies are a function only of the difference between the discrete angles $\theta$, and do not depend on the absolute identity of the patches. In fact, as $N\to\infty$, the energy profile is exactly of the form $\cos\left(\theta_{i}-\theta_{j}\right)$.  This accomplishes the mapping of the buckling transition to a six-state clock model.

\section{Derivation of Coulomb gas}
\label{app:coulombGas}

Recall from the main text that the low energy topological defects in our system are vortices, composite dislocations and double dislocations. Also, remember that each topological defect carries a definite total charge $Q$, and these charges interact via the stiffnesses $K_m$ and $K_l$. Bearing in mind that each class of topological defects has a fugacity, the Coulomb Gas Hamiltonian without the clock term is given by:
\bea
-\beta H_{\mathrm{no\thinspace clock}}&=&\sum_{i<j}\left(K_{m}q_{i}q_{j}+K_{l}\mathbf{b}_{i}\mathbf{\cdot b}_{j}\right)\ln\frac{\left|\mathbf{r}_{i}\mathbf{-}\mathbf{r}_{j}\right|}{a}\nn\\
&\,&+N_{\vor}\ln y_{\vor}+N_{\cd}\ln y_{\cd}+N_{\dd}\ln y_{\dd}
\eea
where $N_{\vor}$, $N_{\cd}$, and $N_{\dd}$ are the total number of each class of defects.
The grand-canonical partition function is:
\bea
Z=\sum_{N_{\vor},N_{\cd},N_{\dd}}\int\prod_{i=1}^{N_{\vor}+N_{\cd}+N_{\dd}}\frac{d^{2}\mathbf{r}_{i}}{a^{2}}e^{-\beta H_{\mathrm{no\thinspace clock}}}
\eea
along with the constraint $\sum_{i}\mathbf{Q}_{i}=0$.

To incorporate the additional clock term into the model, we first replace the cosine function $h_6\cos(6\theta)$ in the original Hamiltonian \eqref{eq:clockHamiltonian} by another function of the same period $g\left(\theta\right)=\ln\left(\sum_{n=-\infty}^{\infty}e^{6i\theta n+n^2\ln\left(y_{\clock}\right)}\right)$, with $y_{\clock}\equiv\frac{h_{6}}{2}$, which makes the problem more tractable \cite{villain1975} . We are allowed to make the replacement since higher Fourier modes do not affect the phase diagram as can be confirmed by RG analysis. By checking two limits of $g\left(\theta\right)$, we validate that it has a similar behaviour to the cosine function: If $y_6<<1$, $g\left(\theta\right)=2y_{6}\cos\left(6\theta\right)+O\left(y_{\clock}^{2}\right)$. If $y_{\clock}\rightarrow1$, the argument of $g\left(\theta\right)$ becomes a sum of delta functions, and all spins relax into one of six angles.

By integrating over the regular part of the $\theta$-variables, the clock term itself can be mapped into a Coulomb gas. The $n$-variables play the role of the charges, and $y_{\clock}$ is analogous to the fugacity of the charges, hence we call it a clock fugacity. In the absence of dislocations, this analogy can be pushed further to show a duality relation between the clock term and the vortices \cite{jose1977}.

After adding the clock term, the resulting Coulomb gas Hamiltonian is,
\bea
-\beta H&=&\sum_{i<j}\left(K_{m}q_{i}q_{j}+K_{l}\mathbf{b}_{i}\mathbf{\cdot b}_{j}
+\frac{6^{2}}{K_{m}}n_{i}n_{j}\right)\ln\frac{\left|\mathbf{r}_{i}-\mathbf{r}_{j}\right|}{a}\nn\\
&\,&+N_{\vor}\ln y_{\vor}+N_{\cd}\ln y_{\cd}+N_{\dd}\ln y_{\dd}+N_{\clock}\ln y_{\clock} \\
&\,&+6i\sum_{i\neq j}q_{i}n_{j}\arctan\left(\frac{y_{i}-y_{j}}{x_{i}-x_{j}}\right)\nn
\eea
where the variables $n_{i}$ take integer values.

The imaginary term in the Hamiltonian originates from the singular part of the $\theta$-variables.  It can be interpreted as an Aharonov-Bohm phase that a topological defect feels in the presence of a dual charge $6n_j$. From renormalization-group point of view, this term is only important in situations where both the clock fugacity and the magnetic vortices are simultaneously relevant, or in direct phase transitions between a phase where the clock term is relevant and a phase where the magnetic vortices are relevant. We ignore this term in the rest of the derivation and find self-consistently that in the phase diagram \ref{fig:phaseDiagClock} there are no such phases where both the clock fugacity and the magnetic vortices are relevant. The only phase transition which may be in principle affected by the term we ignored is EH. However, this transition is a runaway flow and therefore is not expected to be affected by this term.

The resulting partition function is,
\bea
Z=\sum_{N_{\vor},N_{\cd},N_{\dd},N_{\clock}}\int\prod_{i=1}^{N_{\vor}+N_{\cd}+N_{\dd}+N_{\clock}}\frac{d^{2}\mathbf{r}_{i}}{a^{2}}e^{-\beta H}
\eea
The allowed configurations of the partition function are restricted by an additional constraint, $\sum_{i}n_{i}=0$.

\section{Disclinations}
\label{SI:disclinations}

\begin{figure}
\centering
{\begin{overpic}[width=0.23\textwidth]{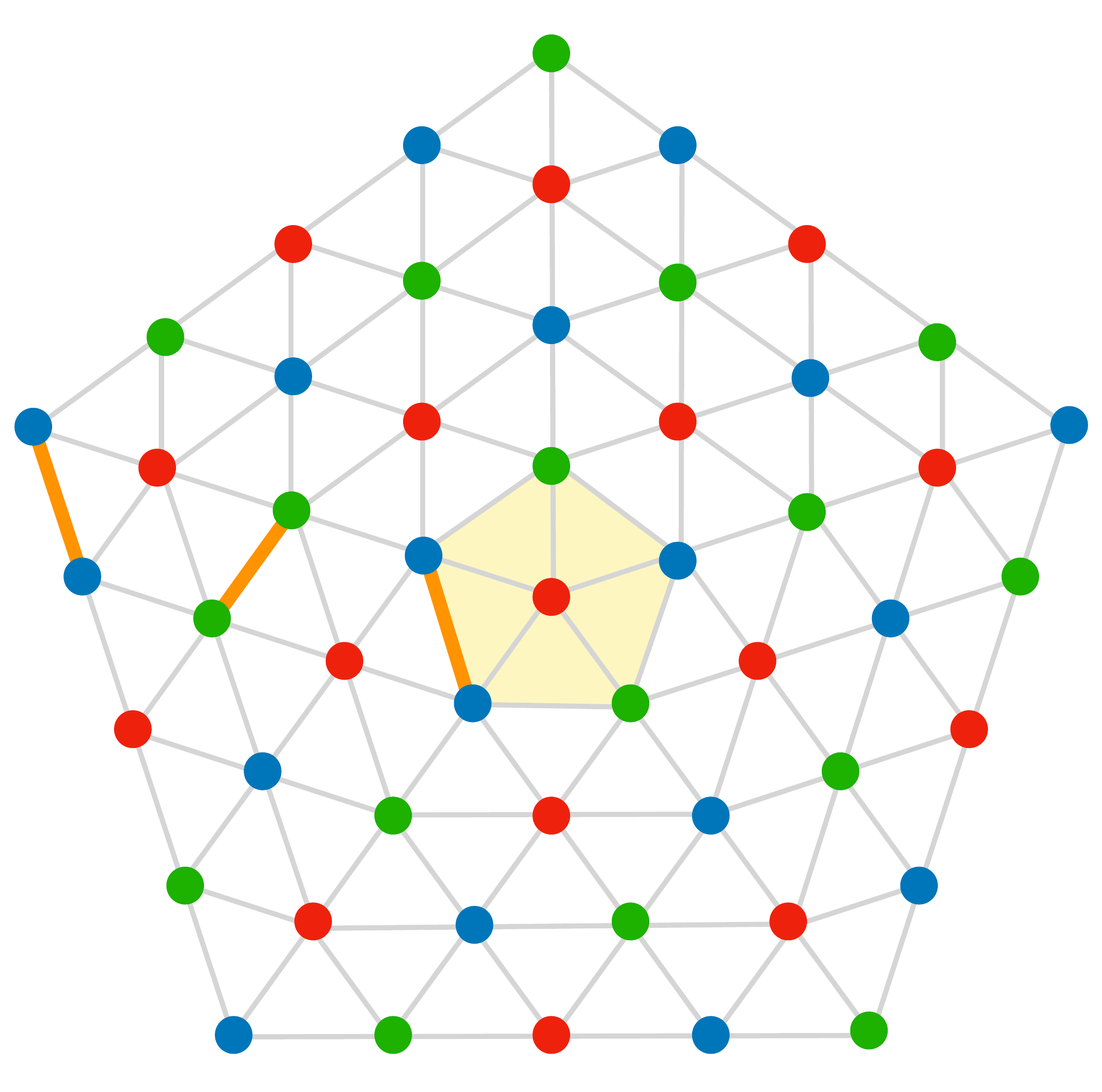}
  \put(0,90){(a)}
 \end{overpic}}
\hfill
\begin{overpic}[width = 0.23\textwidth]{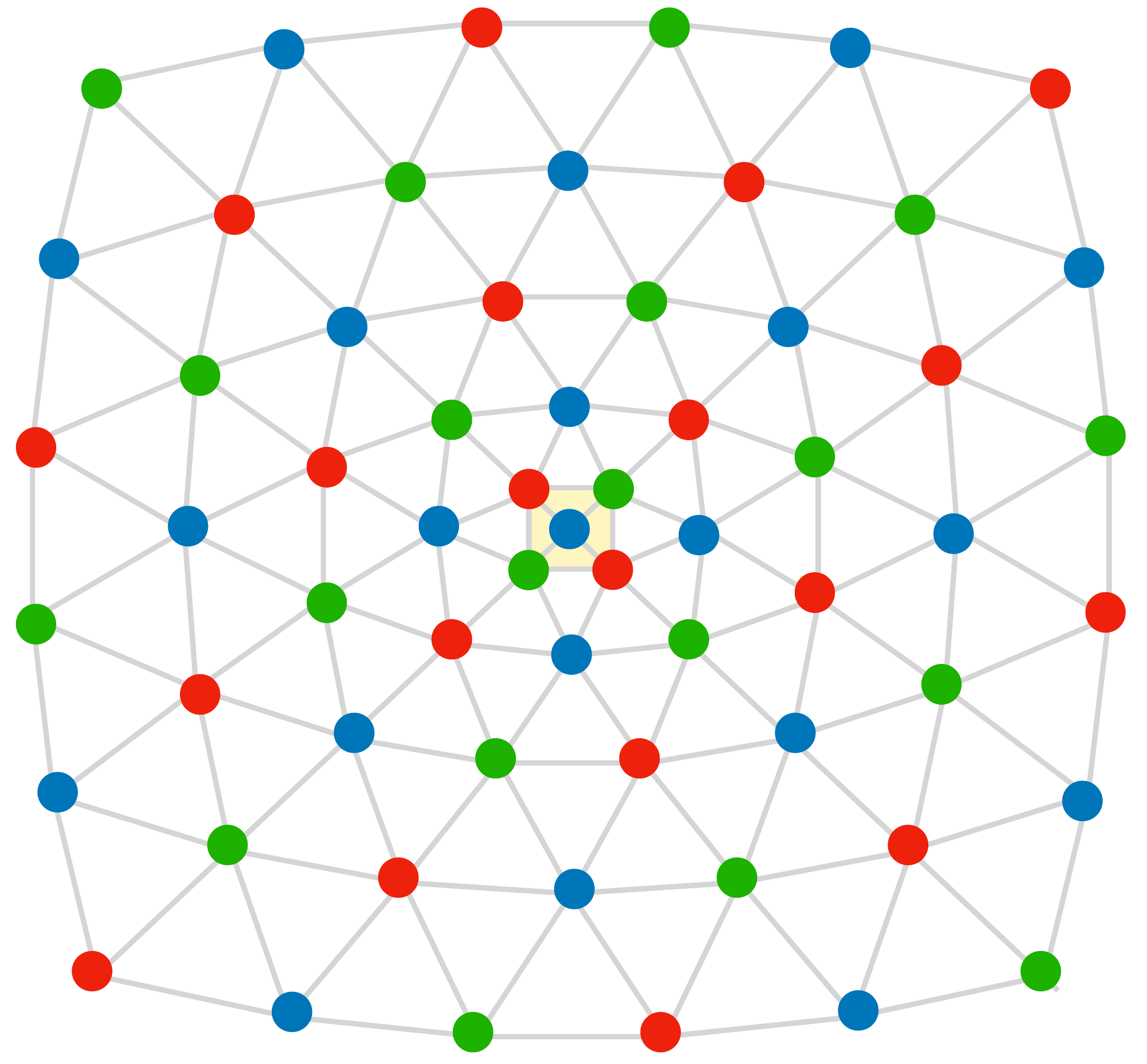}
  \put(-13,90){(b)}
\end{overpic}
\caption{(a) A $\pi/3$ disclination, characterized by a site with five neighbors, generates a domain wall of aligned spins (orange lines).  Unlike the dislocation, adding a fractional vortex does not remove the domain wall. (b) A charge $2\pi/3$ (double) disclination is characterized by a site with four neighbors.  As shown, it does not disrupt tripartite order.  This is also true for a charge $-2\pi/3$ disclination.}
\label{fig:disclinations}
\end{figure}

A disclination is a topological defect of the orientational order of a crystal lattice.  It is characterized by an angular sector of the lattice that is removed (or added) to the lattice.  For instance, the disclination in Fig.~\ref{fig:disclinations}(a) is obtained by removing from the triangular lattice a $\pi/3$ sector emanating from a lattice site.  This site then has five neighbors, one fewer than in the original lattice, and the lattice has five-fold rotational symmetry about this point.  Similarly, in a $-\pi/3$ disclination (not shown), the central site has an extra neighbor, and the lattice has seven-fold symmetry.

As shown in Fig.~\ref{fig:disclinations}(a), in a magnetic system, the $\pi/3$ disclination gives rise to a domain wall in the magnetic ordering.  The domain wall involves sites that are aligned with each other, thus frustrating the antiferromagnetic interactions.  In order to fix this, one would need to exchange the blue and green sites on one side of the domain wall, while leaving the red sites untouched. Hence, we call such a domain wall a transposition-type wall, and label it using cycle notation by $(\blue{\rm B}\ {\rm\green G})$.  In particular, a transposition wall cannot be undone by binding third of a magnetic vortex, which instead yields a cyclic permutation of blue, red, and green,  $(\blue{\rm B}\ {\rm\red R}\ \green{\rm G})$. Hence, within the magnetic phase, the $\pi/3$ disclinations cost an energy that is linear in system size.

As discussed in the main text, a dislocation binds a third of a magnetic vortex.  Since a dislocation is composed of a bound pair of $\pi/3$ and $-\pi/3$ disclinations, following Ref. \onlinecite{gopalakrishnan2013}, it is interesting to ask what happens to the fractional vortex when the disclinations dissociate.   Attached to each disclination, by itself, is a domain wall of transposition-type.  One can then construct a dislocation as a bound state of a $\pi/3$ disclination with a $(\blue{\rm B}\  \red{\rm R})$ domain wall, and a $-\pi/3$ disclination with a  $(\blue{\rm B}\ \green{\rm G})$ domain wall.

One can picture these domain walls as strings that emanate from each disclination.  Then, the lowest energy configuration corresponds to having one string, say of type $(\blue{\rm B}\ \red{\rm R})$, joining the two disclinations.  When the string reaches the second disclination, it merges with the second string.  The merged string is also a domain wall, whose type is the product of the two transpositions, $(\blue{\rm B}\ \red{\rm R})(\blue{\rm B}\ \green{\rm G})=(\blue{\rm B}\ \green{\rm G}\  \red{\rm R})$, that is, a cyclic permutation, which is removed by binding a third of a magnetic vortex.  Hence, as the disclinations are pulled apart, this creates a domain wall of transposition-type between them.  This costs string tension that is linear in the distance between the disclinations, meaning that the disclinations are confined in pairs in the magnetic phase.

On the other hand,  a $2\pi/3$ ({\em i.e.} double) disclination does not frustrate tripartite order.  This can be seen by decomposing it into two identical disclinations, each with a transposition-type domain wall.  Since transpositions square to the identity, these domain walls cancel out.  This can also be seen explicitly, as shown in Fig.~\ref{fig:disclinations}(b).  Note that there are other double disclinations, of plaquette type, which are incompatible with the magnetic order and are forbidden in the magnetic phase\cite{gopalakrishnan2013}.

A double dislocation is a bound state of two double disclinations, as shown in Fig.~\ref{fig:doubleDislocation}.  Hence, it is not surprising that the double dislocation also does not disrupt magnetic ordering.  Of course, one can also understand this directly, from the fact that the Burgers' vector connects sites on the same sublattice of the tripartite order.

\section{Generalization to closely related systems}
\label{app:generalizations}

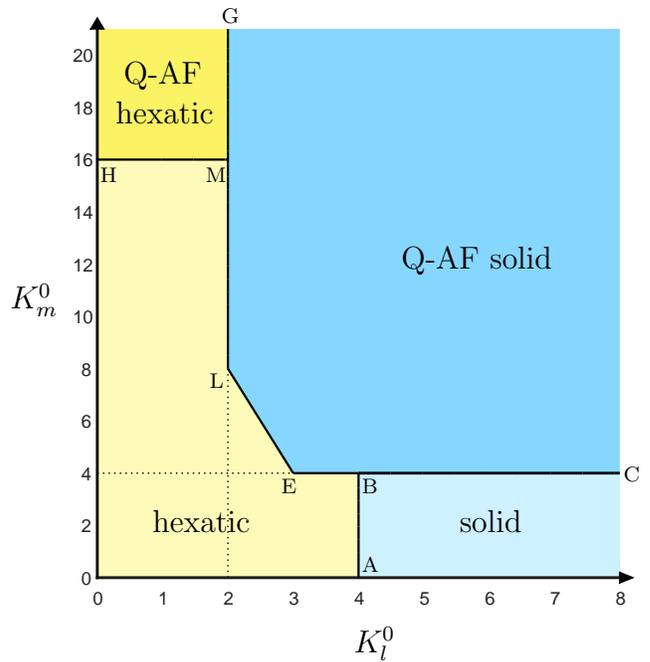
\begin{figure}
\centering
\begin{overpic}[width = 0.5\textwidth]{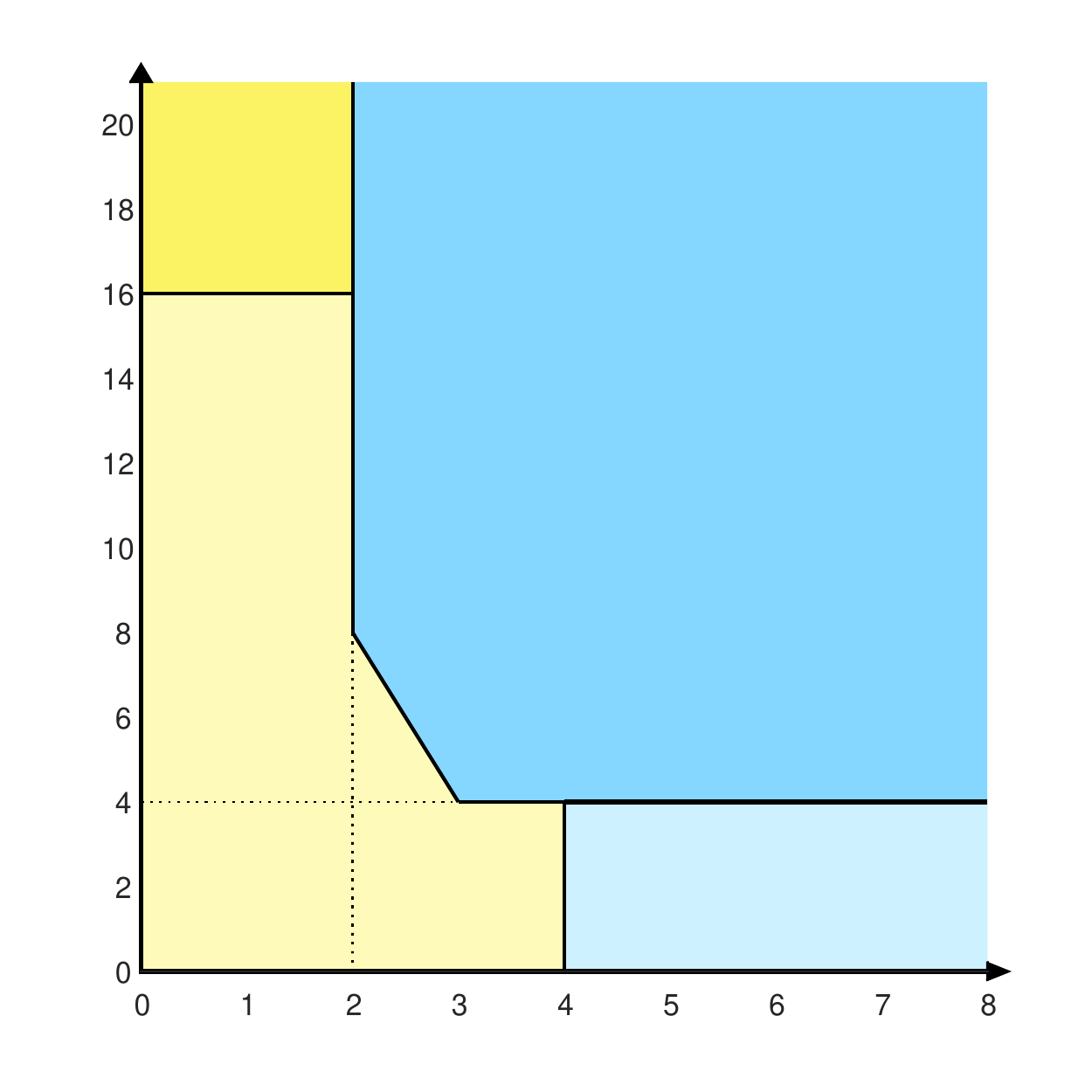}
  \put(0, 130){\large $K_m^0$}
  \put(130, 0){\large $K_l^0$}
  \put(170, 45.5){\large solid}
  \put(148, 145){\large Q-AF solid}
  \put(53.9, 45.5){\large hexatic}
  \put(43, 215){\large Q-AF}
  \put(40, 200){\large hexatic}
  \put(133, 30.5){\footnotesize A}
  \put(133, 60){\footnotesize B}
  \put(232, 64.5){\footnotesize C}
  \put(102.5, 60){\footnotesize E}
  \put(75.5, 100){\footnotesize L}
  \put(74, 178){\footnotesize M}
  \put(80, 238){\footnotesize G}
  \put(34, 178){\footnotesize H}
\end{overpic}
\caption{Phase diagram for the $XY$ antiferromagnet on a square lattice. The phase transitions are: KT transition driven by unbinding of dislocations (AB), KT transition of magnetic vortices (BC), cascaded transition (BE), KT transition of composite dislocations (EL), cascaded transition (LM), KT transition of double dislocations (MG) and KT transition of half vortices (MH).
}
\label{fig:phaseDiagSquareLattice}
\end{figure}

This section illustrates how our analysis generalizes to other closely related systems. We consider two cases: the $XY$ antiferromagnet and the four-state clock model, both on a square lattice.

For the $XY$ antiferromagnet the flow equations are:
\begin{align}
\frac{dK_{m}}{d\ell}&=-2\pi^{2}K_{m}^{2}\left(y_{\vor}^{2}+\frac{1}{2}y_{\cd}^{2}\right)\,,\label{eq:dKmdl}\\
\frac{dK_{l}}{d\ell}&=-2\pi^{2}K_{l}^{2}\left(y_{\cd}^{2}+2y_{\dd}^{2}\right)\,,\\
\frac{dy_{\vor}}{d\ell}&=\left(2-\frac{K_{m}}{2}\right)y_{\vor}\,,\\
\frac{dy_{\cd}}{d\ell}&=\left(2-\frac{K_{l}}{2}-\frac{K_{m}}{8}\right)y_{\cd}\,,\\
\frac{dy_{\dd}}{d\ell}&=\left(2-K_{l}\right)y_{\dd}\,, \label{eq:RGEqationsXyAf}
\end{align}
The equations differ from Eq.~\eqref{eq:RGEqations} by numerical factors due to the square geometry of the lattice (a dislocation binds a half of a vortex). The clock term is absent because the model is $U(1)$-symmetric.

Figure~\ref{fig:phaseDiagSquareLattice} shows the phase diagram for the $XY$ antiferromagnet. True long-range order is not possible in this case, but one finds hexatic Q-AF and liquid Q-AF phases for $K_m\ll K_l$. In the absence of a clock term, all phase transitions become straight lines (assuming infinitesimal bare values for the defect fugacities). The diagonal line EL is given by $K_l+\frac{K_m}{4}=4$. Double dislocations stay as magnetic-free lattice defects, and the vertical asymptote shifts from $K_l=4/3$ to $K_l=2$.  An interesting property of Fig.~\ref{fig:phaseDiagSquareLattice} is the symmetry between lattice and magnetism; the phase diagram is symmetric under the exchange of $K_l\leftrightarrow \frac{K_m}{4}$.

\begin{figure}
\centering
\begin{overpic}[width = 0.5\textwidth]{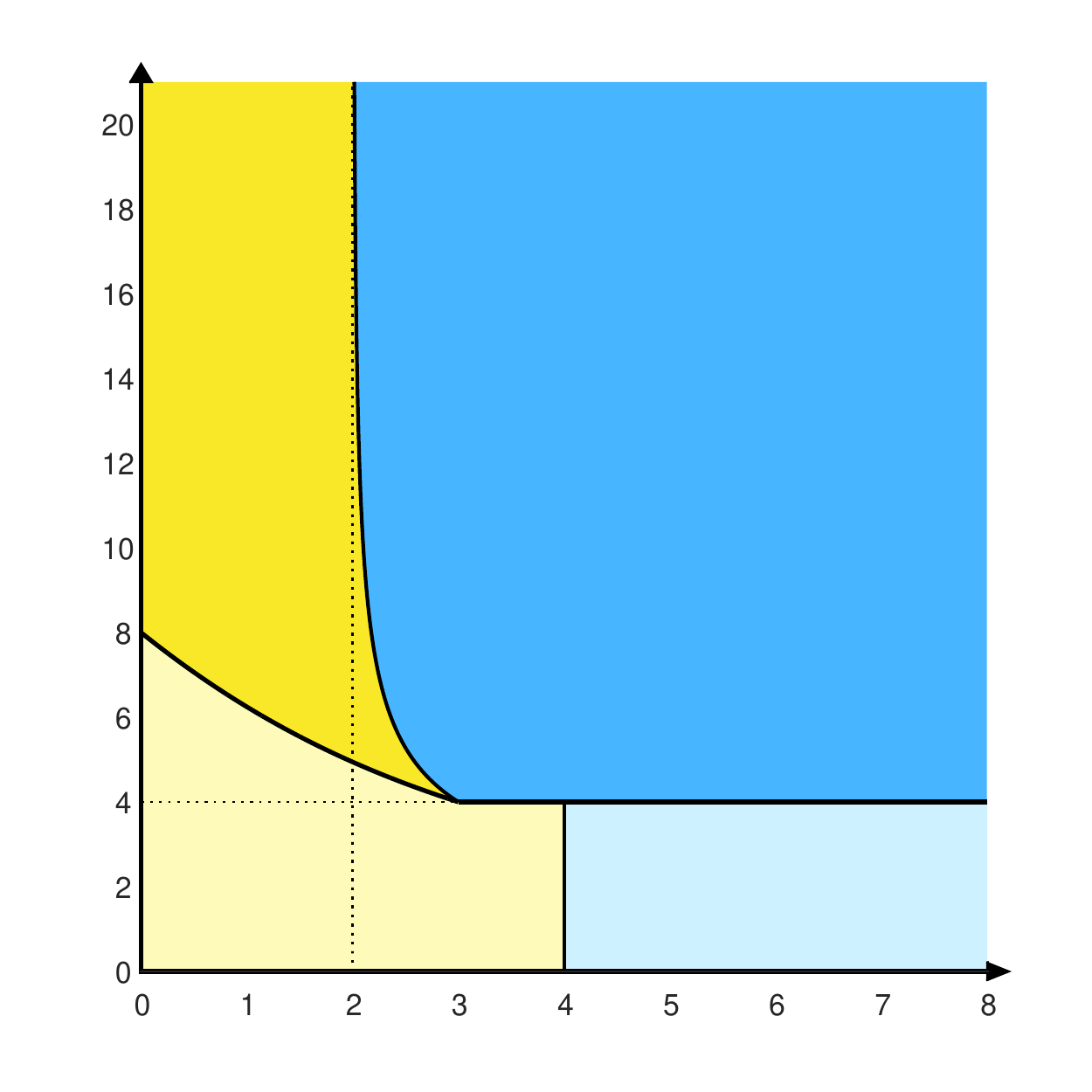}
  \put(0, 130){\large $K_m^0$}
  \put(130, 0){\large $K_l^0$}
  \put(168, 45.5){\large solid}
  \put(150, 145){\large AF solid}
  \put(53.9, 45.5){\large hexatic}
  \put(48, 160){\large AF}
  \put(40, 145){\large hexatic}
  \put(133, 30.5){\footnotesize A}
  \put(133, 60){\footnotesize B}
  \put(232, 64.5){\footnotesize C}
  \put(102.5, 60){\footnotesize E}
  \put(80, 238){\footnotesize G}
  \put(34, 109  ){\footnotesize H}
\end{overpic}
\caption{Phase diagram for the four-state clock model on a square lattice. The phase transitions are similar in nature to those in Fig.~\ref{fig:phaseDiagClock}.}
\label{fig:phaseDiagSquareLatticeClock}
\end{figure}
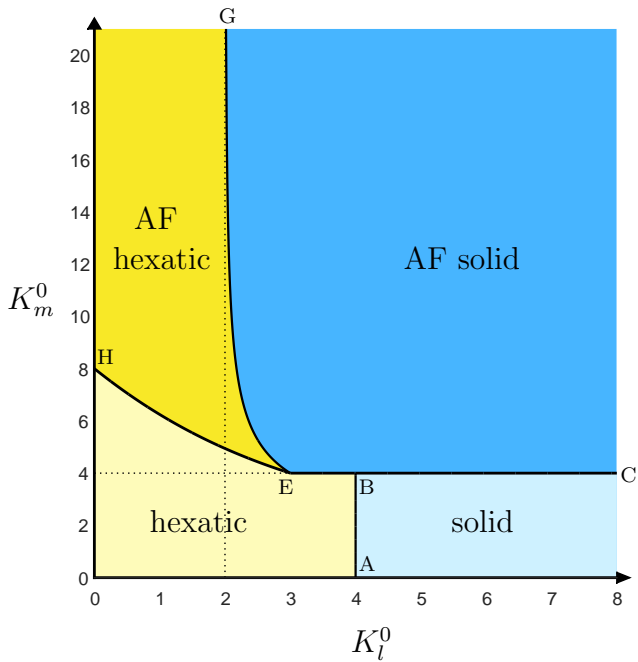

Figure~\ref{fig:phaseDiagSquareLatticeClock} shows the phase diagram for the four-state clock model. The flow equations are similar to the $XY$ antiferromagnet with an extra clock term:  Eq. (\ref{eq:dKmdl}) for $dK_{m}/d\ell$ gets an additional contribution $+2\pi^{2}y_{\clock}^{2}$, and there is aditional equation for the clock fugacity $\frac{dy_{\clock}}{d\ell}=\left(2-\frac{8}{K_{m}}\right)y_{\clock}$.  A comparison of the two phase diagrams for a square lattice, Figs.~\ref{fig:phaseDiagSquareLattice} and \ref{fig:phaseDiagSquareLatticeClock}, highlights the stabilizing effect of the clock term on the formation of the AF-hexatic.

Relating Fig.~\ref{fig:phaseDiagSquareLatticeClock} to the phase diagram for the six state model on a triangular lattice, Fig.~\ref{fig:phaseDiagClock}, we see that, for the square lattice, the vortex unbinding transition and the clock transition merge into a single continuous transition \cite{jose1977} that occurs at $K_m=4$. The points $D$ and $E$ in Fig.~\ref{fig:phaseDiagClock} are identified, and there is no Q-AF solid phase. The hexatic-AF hexatic transition line crossing with the vertical axis is shifted from $K_m=18$ to $K_m=8$. Notice that in this case the AF hexatic can be reached at a significantly reduced value of $K_m/K_l$, since the point $E$ in is now located at $(K_l,K_m)=(3,4)$.

A generalization to the $XY$ model on the triangular lattice is also possible.  However, this requires special care of the chirality, which is known to appear at a separate transition from the $U(1)$ symmetry breaking \cite{miyashita1984nature,kawamura1998universality}.

\bibliography{AFhexaticBib}
\end{document}